\documentclass[showpacs,preprintnumbers,amsmath,amssymb,aps,twocolumn]{revtex4}
\usepackage{graphicx}

\newcommand{\be}{\begin{equation}}
\newcommand{\ee}{\end{equation}}
\newcommand{\ba}{\begin{eqnarray}}
\newcommand{\ea}{\end{eqnarray}}
\newcommand{\bd}{\begin{description}}
\newcommand{\ed}{\end{description}}

\def\di{\displaystyle}

\def\bg{\begin{eqnarray}\begin{array}{rcl}\displaystyle}
\def\eg{\end{array} &\di &\di \end{eqnarray}}
\def\bm#1{\begin{eqnarray}\begin{array}{#1}\di}
\def\bmo#1{\begin{eqnarray*}\begin{array}{#1}\di}
\def\bml#1#2{\begin{eqnarray}\begin{array}{#1}\label{#2}\di}
\def\bgo{\begin{eqnarray*}\begin{array}{rcl}\displaystyle}
\def\ego{\end{array} &\di &\di \nonumber \end{eqnarray*}}

\def\btensor#1#2{\renew\left#1\begin{array}{#2}\di}
\def\brtensor#1#2#3{\ren#3\left#1\begin{array}{#2}}
\def\botensor#1#2{\renew\left#1\begin{array}{#2}}
\def\etensor#1{\end{array}\right#1}

\def\d{{d}}

\def\s0#1#2{\mbox{\small{$ \frac{#1}{#2} $}}}
\def\0#1#2{\frac{#1}{#2}}

\def\bd{{\bf\d}}

\def\CH{{\mathcal H}}

\date{\today}

\def\ren#1{\renewcommand{\arraystretch}{#1}}

\def\renew{\renewcommand{\arraystretch}{1}}

\begin{document}

\title{Corrections to Universal Fluctuations in Correlated
Systems:\\[1ex] the 2D XY-model\\[1ex] }
\author{G. Mack}
\email{gerhard.mack@desy.de} \affiliation{ II. Institut
f\"ur
Theoretische Physik, Universit\"at Hamburg, D-22761
Hamburg,
Luruper Chaussee 149, Germany,}
\author{G. Palma}
\email{gpalma@lauca.usach.cl}
\author{L. Vergara}
\email{lvergara@lauca.usach.cl,}
\affiliation{\mbox{Departamento de F\'{\i}sica,
Universidad de Santi\-ago de Chile,}\\
Casilla 307, Santi\-ago 2, Chile. }

\begin{abstract}
Generalized universality, as recently proposed, postulates
a
universal non-Gaussian form of the probability density
function
(PDF) of certain global observables for a wide class of
highly
correlated systems of finite volume $N$. Studying the 2D
$XY$-model, we link its validity to renormalization group
properties. It would be valid if there were a single
dimension $0$
operator, but the actual existence of several such
operators leads
to $T$-dependent corrections. The PDF is the Fourier
transform of
the partition function $Z(q)$ of an auxiliary theory which
differs
by a dimension $0$ perturbation with a very small
imaginary
coefficient $iq/N$ from a theory which is asymptotically
free in
the infrared. We compute the PDF from a systematic loop
expansion
of $\ln Z(q)$.
\end{abstract}

\maketitle

Trying to understand universality properties of critical
systems
near a second order phase transition was extremely
fruitful for
the development of physics. It lead to the development of
Wilsons
renormalization group (RG)~\cite{Wilson} which found
applications
ranging from solid state physics to elementary particle
theory.
Universality classes in the RG-sense depend on the
dimension of
the system and on the symmetry properties of the order
parameter.

Recently a generalized universality has been
proposed~\cite{Bram1}, which
goes far beyond the known picture. It is therefore of
great general interest
to understand the conditions for its validity. Generalized
universality is
supposed to hold true for systems sharing the properties
of strong
correlations, finite volume and self similarity, no matter
whether they have
the same symmetries and dimensions nor whether they
correspond to
equilibrium or non-equilibrium systems. Such new
universality is expressed
in a non-Gaussian universal curve these systems share when
the PDF of some
global quantity of each system is plotted (universal
fluctuations). Studying
the 2D XY-model as an example, we link its validity to
RG-properties of the
system, viz. the existence of dimension 0 perturbations of
a system which is
asymptotically free in the infrared. It would be exactly
true if there were
one single dimension 0 operator. But the actual existence
of several
dimension 0 operators in the model leads to T-dependent
corrections to the
supposedly universal curve which do not go away when the
volume $N\mapsto \infty$. If there were no dimension 0
operator, the PDF
would be Gaussian.

Let us recall some previous results. The supposed
universal curve for the
PDF, which we call BHP, is non-Gaussian in spite of the
result that a naive
application of the central limit theorem would have given;
this possibility
has been attributed~\cite{Bram1} to the anomalous
statistical property of
finite size critical systems that, whatever their size is,
cannot be divided
into mesoscopic regions that are statistically
independent, a necessary
condition for the central limit theorem to apply. In
Ref.~\cite{Bram2} it
was shown that a steady state in a closed turbulent flow
experiment and a
finite volume magnetic model, the 2D XY-model in a finite
square lattice,
had the same universal non-Gaussian probability
distribution function for
the fluctuations of a global quantity. These are two
examples of systems,
among others
(see~\cite{BTW},~\cite{SNP},~\cite{PSR},~\cite{ECG}), that
exhibit such data collapse.


In~\cite{Golden} it was argued that in finite-size systems
that are in the
critical regime standard scaling is a sufficient condition
to exhibit data
collapse. Nevertheless, as our results show, when one
computes instead of
the PDF more sensitive quantities (from the numerical
point of view), such
as its moments (skewness and kurtosis), temperature
dependence shows up.
This occurs within the range of temperatures of the
expected data collapse.


A functional form for the BHP universal curve was suggested
in~\cite{Bram1} and it was also argued that this universal curve
should be independent of both size and temperature of the system.
This result was based on an earlier computation of the PDF of the
magnetization of the 2D XY-model in the harmonic (=spin-wave)
approximation (the 2D HXY-model)~\cite{Archam}. In this reference
the PDF of the magnetization was carried out by resumming its
moment series
up to leading order in 
$N$ (the system 
size), assuming that multiple loop diagrams can be
neglected,
  and finally by performing an inverse Fourier integral.

Nevertheless, more recently~\cite{Palma1} a numerical
study of the full 2D
XY-model was performed, and the results suggested a small
but systematic
dependence of the PDF on the system temperature. This
numerical analysis was
performed by using high precision Monte Carlo simulations,
which amounts to
control the statistical independence of the configurations
used to compute
thermodynamical averages, as this system is affected by
the well known
critical slowing down effect.

In this work we explain how to compute systematically the PDF for
the 2D XY-model within the loop expansion of an auxiliary theory
with an imaginary coupling. It turns out that the 1-loop
approximation agrees with the curve BHP computed in
Ref.~\cite{Bram3} . But higher loop corrections have a different
$T$-dependence. They are not suppressed by powers of $1/N$ and the
$T$-dependence of their contribution to supposedly universal
quantities such as skewness and kurtosis does not disappear for
large $N$. This unexpected breakdown of strict BHP universality is
not due to the corrections to the harmonic approximation to the
Hamiltonian, but comes from the fact that the nonlinear order
parameter (the magnetization) is a sum of several dimension 0
operators which contribute at different order of the loop
expansion.

We will later confirm the known fact ~\cite{Bram3} that
the
corrections to the spin wave approximation will not affect
the
PDF, for temperatures well below the
Berezinskii-Kosterlitz-Thouless critical temperature
$T_{BKT}$ and
for large $N$, apart from a renormalization of
temperature.
Therefore the PDF for the XY and the HXY-model have the
same large
$N$ limit except for a renomalization of temperature, but
neither
of them obeys strict BHP-universality.

  Because vortices are neglected in perturbation theory,
results are only valid well below $T_c$

Let us consider a classical statistical mechanical system
which
lives on a lattice with $N$ sites $x$, to which variables
$\phi_x$
are attached, and with a Boltzmann weight $e^{-\beta
\mathcal{H}(\phi)}D\phi$, where $D\phi=\prod_x d\phi_x$
and
$d\phi_x$ is a measure on the space in which $ \phi_x$
takes its
values. Let $A(\phi)$ be a global quantity of the form
\begin{equation}
A(\phi) = \frac 1 N \sum_x \Phi(\phi_x) , \nonumber
\end{equation}
where $\Phi(\phi_x)$ is some function of $\phi_x$. The
magnetization per site in a ferromagnet is an example of
such a
global variable.

The PDF is defined as
\begin{equation}
P(M)= \frac 1 Z_0 \int D\phi e^{-\beta
\mathcal{H}(\phi)}\delta
(A(\phi)-M)
\end{equation}
with partition function $Z_0= \int D\phi e^{-\beta
\mathcal{H}(\phi)}$. The mean $\langle M\rangle$ and mean
square
fluctuation $\sigma^2=\langle(M- \langle M\rangle
)^2\rangle$ of
$A$ are computable from the PDF as $\langle M\rangle =\int
M \
P(M)dM $ and $\sigma^2 = \int (M-\langle M\rangle )^2
P(M)dM$,
respectively.

Inserting the Fourier representation of the
$\delta$-function, we have

\begin{equation}
P(M)=\int_{-\infty}^{\infty } \frac {dq}{2\pi}\,
e^{-iq(M-\langle
M\rangle )} Z(q), \label{3}
\end{equation}
where
\begin{equation}
Z(q) = \frac {e^{-iq\langle M\rangle}} {Z_0} \int D\phi
exp \,
\left\{-\beta \mathcal{H}(\phi) + \frac {iq}{N}\sum_x
\Phi(\phi_x)\right\} . \label{4}
\end{equation}

The hypothesis of BHP universality says that the PDF,
considered
as a function
\[
P(M)= F\left( \frac{M-\langle M\rangle }{\sigma}\right)
\]
of $(M-\langle M\rangle )/\sigma $, is a universal
function $F$,
the same for many strongly correlated critical systems,
independent of temperature $T=\beta^{-1}$ and of system
size $N$,
provided $N$ is large enough.

We see that the PDF is the Fourier transform of the
``partition
function'' $Z(q)$, normalized to $Z(0)=1$, of an auxiliary
theory
whose Hamiltonian differs from that of the original theory
by a
perturbation with an imaginary coupling $iq/N$ which is
small of
order $1/\mbox{volume}=N^{-1}$. On finite systems, $Z(q)$
will
typically be a holomorphic function of $q$ in a small
neighborhood
of zero, and $Z(q)$ is a true partition function for
positive
imaginary $q$.
We choose to extract the factor $\exp (iq\langle M
\rangle) $ from
the partition function so that its derivative
$Z^\prime(0)=0$.

The hypothesis of BHP-universality asserts that for the
above mentioned
systems, the ``partition function'' $Z(q)$ is a universal
non-Gaussian
function of $x=-\sigma q$. %
We consider the magnetization per site of the
2-dimensional
XY-model. $ \phi_x $ can be considered as real variables,
and the
Hamiltonian is

\begin{equation}
\mathcal{H}(\phi)=-J \sum_{<x,y>} \cos (\phi_x-\phi_y).
\end{equation}
Without loss of generality, the coupling constant $J>0$
can be set to $1$.
The sum runs over pairs of nearest neighbor sites $x,y$ on
a square lattice
of side length $L=\sqrt{N}$. The Hamiltonian is invariant
under a global
rotation of the spins $(\cos \phi_x , \sin \phi_x)$ and
under
translations $\phi_x \mapsto \phi_x+2\pi n_x$, $n_x$
integer.
By a global shift of the variables $\phi_x$ one can
achieve
$\sum_x \phi_x = 0$. The magnetization per site is then
given by
\begin{equation}
\Phi(\phi_x)= \cos \phi_x .
\end{equation}
The density $\cos\phi_x$ is a sum of dimension 0
perturbations which are
eigenmodes of the linearized renormalization group. The
existence of a
dimension 0 perturbation explains why the $q$-dependent
perturbation can
make a contribution of order $N^0$ to $\ln Z(q)$, in each
order $q^n$,
despite the small coefficient $\propto N^{-1}.$ The
existence of several (in
fact infinitely many) dimension 0 perturbations explains
why there is no
strict BHP-universality. If the perturbation had dimension
$>0$, successive
orders in $q$ would be suppressed by negative powers of
$N$, so that the
term $\propto q^2$ would dominate after the term $\propto
q$ is subtracted
from $\ln Z(q)$, and the PDF would be Gaussian.

The existence of several dimension 0 operators is unusual
even in
two space dimensions. But for the 2D XY-model it follows
from
features of free field theory because the (known) accuracy
of the
harmonic approximation to the Hamiltonian implies that we
consider
RG-transformations in the vicinity of a Gaussian fix
point.

In the spin wave approximation, the Hamiltonian is
approximated by
\begin{equation}
\mathcal{H}_{SW}(\phi) = \frac 1 2 \sum_x \sum_{\mu=1}^2
|\nabla_\mu \phi (x)|^2,
\end{equation}
where $\nabla_\mu $ is the finite difference derivative.
This is
the Hamiltonian of the HXY-model.

Since $\beta=1/T$ appears as a factor in front of
$\mathcal{H}$,
the logarithm of $Z(q)$ can be computed as a power series
in $T$,
for fixed $qT/N$. This is called the loop expansion since
the
parameter $T$ plays the role of $\hbar$ in field
theory~\cite{Palma2}. The expansion up to $n$ loops
includes terms
up to order $T^{n-1}$.

On finite lattices, $\ln Z(q)$ may be expanded in a power
series in $q$.
\begin{equation}
iq\langle M \rangle + \ln Z(q) = iq + \sum_{n \geq 1}
(-iqT)^n
\alpha_n(T) \label{expand}
\end{equation}
The coefficients $\alpha_n$ also depend on system size
$N$. By definition,
$Z(q)$ is normalized to $1$ and obeys $Z^\prime(0)=0$,
hence
\begin{equation}
1-\langle M \rangle = T\alpha_1(T)
\end{equation}
and the mean square fluctuation $\sigma^2 = -
\frac{d^2}{d^2q^2}\ln Z(q=0)$ equals
\begin{equation}
\sigma^2 = 2T^2 \alpha_2(T).
\end{equation}
To obtain the PDF with normalized first and second
moments, one considers $\ln Z(q)$ as a function of
$x=-q\sigma$. It has the
form
\[
-iq(M-\langle M\rangle) + \ln Z(q) = ix\frac {M-\langle M
\rangle}{\sigma } - \frac {x^2}{2} + \Psi(x),
\]
with $\Psi(x)=O(x^3)$, and the higher moments are obtained
from it
as
\begin{equation}
\langle \left( \frac {M-\langle M\rangle}{\sigma}
\right)^n\rangle
= \left( i \frac d {dx}\right)^n \exp \left( -\frac 1 2
x^2 +
\Psi(x)\right)_{x=0}, \label{moments}
\end{equation}
for $n\geq 3$.
Inserting the power series expansion (\ref{expand}) of
$\ln Z $
into eq.(\ref{3}) results in
\begin{eqnarray}
P(M) &=&\int_{-\infty }^{\infty }\frac{dx}{2\pi \sigma
}\,\exp
\bigg\{ix \frac{(M-\langle M\rangle )}{\sigma
}-\frac{1}{2}x^{2} \nonumber \\
&&+\sum_{n\geq 3}\alpha _{n}(T)\left( \frac{ixT}{\sigma
}\right)
^{n}\bigg\}, \label{masmm}
\end{eqnarray}
The skewness $s=\langle(M-\langle M \rangle )^3\rangle
\sigma^{-3} $ and kurtosis $c = \langle(M-\langle M
\rangle )^4\rangle \sigma^{-4}$ are obtained from
eq.(\ref{moments}) as
\begin{eqnarray}
s &=&-6\left( \frac{T}{\sigma }\right)
^{3}\alpha _{3} \\
c &=&3\left\{ 1+8\left( \frac{T}{\sigma } \right)
^{4}\alpha
_{4}\right\} \label{mus}.
\end{eqnarray}

BHP-universality would require that the coefficients
$\alpha_n(T)(T/\sigma)^n$ of $(ix)^n$ have universal
values, given
by the BHP-formula $g_n(\frac 1 2 g_2)^{-n/2}/2n$
  which we recover
below as the 1-loop approximation.
In particular, these coefficients should be
$T$-independent, and also the skewness and kurtosis should
be $T$-independent. We will compute the coefficients
$\alpha_n$ to 2-loop order.

We extract the quadratic terms
\begin{eqnarray}
\beta \mathcal{H}(\phi) &=& \beta \mathcal{H}_{SW}(\phi) +
V_1(\phi), \\
V_1(\phi) &=& -\beta \sum_x\left(\frac 1 {4!} (\nabla
\phi_x)^4 +
\dots
\right) \\
\frac {iq}{N} \sum_x\cos \phi_x &=& \frac {iq}{N}\sum_x
(1- \frac
1 2 \phi_x^2) - V_2(\phi_x)
\end{eqnarray}
and define the q-dependent propagator
\begin{equation}
\Gamma = (I+\frac {iqT}{N}\,G)^{-1}G , \label{Gamma}
\end{equation}
where
\begin{eqnarray}
G(x)&=& N^{-1}\sum_{\mathbf{k}\neq 0} e^{-ikx}
\tilde{G}(\mathbf{k}), \\
\tilde{G}(\mathbf{k}) &=&\frac 1 {\mathbf{K}^2}, \\
\mathbf{K}^2&=&\sum_{\mu=1,2} \sin^2 (k_\mu /2)
\end{eqnarray}
is the propagator of a massless scalar free field on a
finite 2
dimensional
lattice with the zero mode removed.

Using the familiar device of introducing a source term
$\langle j ,\phi\rangle$, the integral
defining $Z(q)$ can be formally performed~\cite{Symanzik}
with the result
\begin{eqnarray}
Z(q)=\frac 1 Z_0 e^{-iq(\langle M\rangle -1)}
\exp\left\{\frac 1 2
Tr \ln
[T\Gamma]\right\}\cdot \nonumber \\
\cdot \exp\left\{ -V\left[ \frac {\delta}{\delta j}\right]
\right\} \exp \left\{ \frac {T}{2} \langle j, \Gamma j
\rangle\right\}_{j=0} , \label{exactPertResult}
\end{eqnarray}
with $V=V_1+V_2$.

In 1-loop approximation, we obtain from Eq.
(\ref{exactPertResult})
\begin{equation}
-iq(M-\langle M\rangle) + \ln Z(q) = -iq(M-1)+\frac 1 2 Tr
\ln
[\Gamma] \nonumber
\end{equation}
plus $q$-independent terms.
Upon inserting the definition (\ref{Gamma}) of $\Gamma$,
\be Tr \ln \Gamma = \sum_{\mathbf{k}\neq 0}
  \left\{\ln \tilde{G}(\mathbf{k}) -\ln \left( 1 + \frac
{iqT}{N}\tilde{G}(\mathbf{k})\right) \right\}
\label{TrLnGamma}
\ee
the PDF in 1-loop approximation can be computed by
numerical
Fourier transformation according to eq.(\ref{3}). A
similar
computation is also possible when the 2-loop correction is
added;
in harmonic approximation it requires
\be \Gamma (0) = \frac 1 N Tr \Gamma \nonumber = \frac 1 N
\sum_{\mathbf{k}\neq 0} \left( 1 + \frac {iqT}{N}
\tilde{G}(\mathbf{k}\right)^{-1}\tilde{G}(\mathbf{k})
\label{Gamma0} \ee
according to eq.(\ref{masm}) below.

This result of the PDF in 1-loop approximation reproduces
exactly
the previous result~\cite{Bram1},~\cite{Bram3} for the BHP
PDF. In
section IIA of~\cite{Bram3} it is stated ``\emph{It
appears that
the values of multiple loop graphs, like the first one in
Fig.
1c), are zero in the thermodynamic limit}''. But in fact
this is
not true for all graphs, and as a result the BHP PDF
is merely a lowest
order approximation which becomes exact in the limit
$T\mapsto 0$ only.

We will compute the 2-loop contribution below and comment
on the
precise relation between the diagrammatic expansion here
and
in~\cite{Bram3} later on.

Following Ref.~\cite{Bram1} we define the quantities $g_n$
via powers of $G$
as
\begin{equation}
G^n(0)=N^{-1}Tr G^n= N^{n-1}g_n .
\end{equation}
$g_1$ is logarithmically divergent in the thermodynamic
limit,
$g_1 \sim \frac 1 {4\pi} \ln CN$, $C=1.8456$ ~\cite{Bram3}
while
the remaining ones
become $N$-independent for large enough $N$ and in the
thermodynamic limit they read
\begin{eqnarray}
\quad g_2 &=&0.38667\ 10^{-2}, \nonumber \\
\quad g_3= 0.7572\ 10^{-4}&,&\quad g_4 = 1.763\ 10^{-6}. \nonumber
\end{eqnarray}

Expanding the q-dependent propagator in powers of $q$ we
obtain the 1-loop approximation $\alpha^{(1)}_n(T)$ of the
coefficients $\alpha_n(T)$, viz.
\begin{equation}
\alpha^{(1)}_n(T) = \frac {g_n}{2n} \label{alpha1}
\end{equation}
together with the results for the expectation value
$\langle M\rangle$ of
the magnetization and its mean square fluctuation
$\sigma^2$ in 1-loop
approximation
\begin{eqnarray}
\langle M\rangle^{(1)} &=& 1-Tg_1/2 \\
(\sigma^{(1)})^2 &=& T^2 g_2/2 .
\end{eqnarray}
Note that $\sigma$ is proportional to $T$, so that the
factors
$\left( T/\sigma\right)^k$ are $T$-independent in 1-loop
approximation.

Inserting the result (\ref{alpha1}) into eq.(\ref{masmm})
we recover the known formula of BHP for the universal PDF.

Numerically, the skewness $s=\langle (M-\langle M
\rangle)^3\rangle\sigma^{-3}$ and kurtosis $c=\langle
(M-\langle
M\rangle)^4\rangle \sigma^{-4} $ in 1-loop approximation
are
obtained from Eq. (\ref {moments}) as $s=0.8907$ and
$c=4.4150$
for $N\mapsto \infty$.

In 2-loop approximation
there are two corrections, $\alpha_n = \alpha_n^{(1)} +
\alpha_n^{(2harm)} + \alpha_n^{(2anh)}$.
The first correction comes from the $\phi_x^4$
contribution to
$\frac {iq}{N}\cos \phi_x$. Its contribution to
$-iq(M-\langle M\rangle )+\ln Z(q)$ is
\begin{equation}
\frac{iqT^2}{8}[\Gamma (0)]^{2}\ . \label{masm}
\end{equation}
Expanding $\Gamma $ in powers of $q$ gives
\begin{equation}
\alpha_k^{(2harm)} = - \frac T 8 \sum_{m=0}^{k-1}
g_{m+1}g_{k-m} \
. \label{alpha2harm}
\end{equation}
This term is also present in the HXY-model.

The second correction comes from the anharmonic term
$-\beta
(\nabla \phi_x)^4/4!$ in the Hamiltonian. It is absent in
the
HXY-model. It evaluates to
\begin{equation}
\frac{T}{8}N\left( [\Delta \Gamma (0)]^{2}-[\Delta
G(0)]^{2}\right) \ .
\end{equation}
The $q$-independent subtraction $\propto [\Delta
G(0)]^{2}$ restores the normalization $\ln Z(0)=0$.
Expanding in powers of $q$ and including terms of order
$1/N$ results in
\begin{equation}
\alpha_k^{(2anh)} = \frac T 4 \left( g_k + \frac 1 {2N}
\sum_{l=1}^{k-1} g_{k-l}g_l \right) \ . \label{anh}
\end{equation}
We shall argue below that to leading order in $1/N$ this
anharmonic correction
only amounts to a renormalization of temperature which
cannot lead to a violation of BHP-universality.

  The presence
of the $1/N$-corrections may explain why, for finite
lattice sizes, the shape
of the PDF of the full 2D XY-model displays a stronger
temperature
dependence than the spin-wave approximation, as shown
in Ref. \cite{Palma1}.

In 2-loop approximation, to next to leading order in $T$,
the average
magnetization, mean square fluctuation, skewness $s$ and
kurtosis $c$
including anharmonic corrections
come out as
\begin{equation}
\langle M\rangle
=1-\frac{T}{2}g_{1}-\frac{T^{2}}{4}g_{1}+\frac{T^{2}}{8}%
g_{1}^{2}, \label{2loop1}
\end{equation}
\begin{equation}
\sigma ^{2} =T^{2}\frac{g_{2}}{2}\left\{ 1-Tg_{1}+T\
\left( 1+\frac{1}{2N}%
\frac{g_{1}^{2}}{g_{2}}\right) \right\}, \label{2loop2}
\end{equation}
\begin{eqnarray}
s &=&-g_{3}\left( \frac{2}{g_{2}}\right) ^{3/2}\bigg
\{1+T\bigg(-\frac{3}{4}
\frac{g_{2}^{2}}{g_{3}}+\frac{3}{2N}\bigg
[\frac{g_{1}g_{2}}{g_{3}}
\nonumber \\
&&-\frac{g_{1}^{2}}{2g_{2}}\bigg ]\bigg )\bigg \} ,
\label{2loop3}
\end{eqnarray}
\begin{eqnarray}
c &=&3\bigg\{1+\frac{4
g_{4}}{g_{2}^{2}}+\frac{4T}{g_{2}^{2}}\bigg(%
-2g_{2}g_{3}+\frac{1}{N}\bigg[2g_{1}g_{3} \nonumber \\
&&+g_{2}^{2}-\frac{g_{1}^{2}g_{4}}{g_{2}}\bigg]\bigg)\bigg\}\
.
\label{2loop4}
\end{eqnarray}
Using the values of $g_n$ for the appropriate volumes $N$,
skewness and
kurtosis for the $XY$-model evaluate to
\begin{eqnarray}
s &=& -0.891 + 0.1319 \ T \quad (N=\infty)\\
&=& -0.854+0.1555\ T \quad (N=16^2) \label{2loop3num}\\
c &=& 4.415 - 0.470 \ T \quad (N=\infty) \\
&=& 4.328 - 0.504 \ T \quad (N=16^2).\label{2loop4num}
\end{eqnarray}

As stated above, terms suppressed by powers of $N^{-1}$
come from the
anharmonic corrections.

We see that both $s$ and $c$ have a $T$-dependence which
does not
disappear as $N\mapsto \infty$. This disproves strict
BHP-universality.

The values for $N=\infty$ and $N=16^2$ are close to each
other,
yet different enough to suggest that comparison with Monte
Carlo
data at $N=16^2$ will yield a sensitive test of the
accuracy of
the calculation.

In Fig. 1 we show Monte Carlo results, similar to those of
Ref.
\cite{Palma1}, for the skewness and kurtosis in the
temperature
range $T=0.2,...,1.8$
together with the $T=0$ result given by the 1-loop
approximation, and the
2-loop results (\ref{2loop3num}), (\ref{2loop4num}) up to
order $T$.
  The agreement is quite satisfactory.
\begin{figure}[h]
\centering 
\hspace{-2 cm}
\includegraphics[width=.35\textwidth]{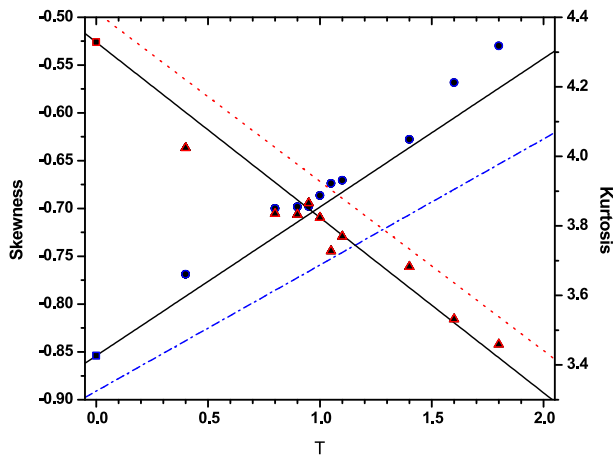}
\vspace{-.2cm} \caption{Skewness and kurtosis. Full
triangles and
circles are Monte Carlo data for a lattice with $N=16^2$
points.
Full lines correspond to the numerical values given in
Eqs.
(\protect\ref{2loop3num},\protect\ref{2loop4num}). The
exact
1-loop results at $T=0$ are shown as full square dots.
Dotted and
dash-dotted lines correspond to the thermodynamic limit
($N=\infty$).} \label{panel}
\end{figure}

In 1-loop approximation, the $(iq/N)\cos
\phi_x$-perturbation is
approximated by $(iq/N)(1-\phi_x^2/2)$. The 2-loop
correction
takes the $\phi_x^4$-term in the Taylor expansion of $\cos
\phi_x$
into account. The
different $T$ and $x$-dependence of the 2-loop correction
to $\ln
Z(q=-x/\sigma)$ can therefore be attributed to the fact
that it
comes from a
different relevant operator of dimension 0.

We will now argue that the effect of the corrections to
the spin wave
approximation on the PDF is merely a renormalization of
temperature, up to
corrections of order $N^{-1}$, viz.
\[
T \mapsto T(1+T/2 + ...)
\]
This comes from the fact that the theory with Hamiltonian
$\mathcal{H}$ is
asymptotically free in the infrared.

The 1-loop approximation is not affected by the
renormalization of
temperature, because it is $T$-independent, and the 2-loop
correction is
also unaffected to order $T$, because the correction would
be of order $T^2$.
Thus, to order $T$,
the skewness and kurtosis in the XY and the
HXY models differ only by corrections proportional to
$1/N$.

Because the coefficient of the $q$-dependent perturbation in
$Z(q)$ is of order $N^{-1}$, the $q$-dependent part of $\ln Z(q)$
is entirely determined by the infrared behavior of the theory, up
to corrections of order $N^{-1}$. Therefore it could be computed
from an effective low energy theory. In principle its Hamiltonian
would have to be computed by a real space renormalization group
(RG) calculation \cite{Wilson}. Since $V_1$ is a sum of irrelevant
perturbations of massless free field theory, their coefficients
get suppressed when the length scale gets enlarged, but they can
give rise to a renormalization of the coefficient $T^{-1}$ of the
marginal $|\nabla \phi_x|^2$ term. This is the aforementioned
renormalization of temperature. In principle there could also be a
renormalization of the coefficients of the relevant perturbation,
i.e. of the coefficient $q$ of $\phi^2$ in 1-loop approximation.
But a multiplicative renormalization of $q$ drops out when the
second moment of the PDF is normalized. New terms in the perturbed
effective action like $ T^3\cos 2\phi_x$ can appear in higher
orders, but they are too small to be detectable.

Note, however, that the perturbative treatment of the
corrections to the
spin wave approximation does not take vortices into
account. Its results are
therefore only valid well below $T_c$.

To verify the RG-argument, we inspect the anharmonic
corrections
  (\ref{anh}). Addition of $T g_k/4$ to $\alpha_k^{(1)}=
g_k/2k$ corrects the 1-loop
  result by the substitution
\begin{equation}
g_k \mapsto g_k (1+kT/2)
\end{equation}
to first order in $T$. Clearly, this has no effect on
$g_k\sigma^{-k}$ to
order $T$, and therefore the normalized PDF remains
unchanged. %

Next we discuss the precise relation between our loop
expansion and the
diagrammatic expansion of Ref. \cite{Bram3}, and
hyperscaling.

One may envisage generalization of the kinetic term \be \frac
\beta 2 \sum_x \sum_{\mu}|\nabla_\mu \phi_x|^2 = - \frac \beta 2
\langle \phi , \Delta \phi \rangle = \frac 1 {2T} \langle \phi ,
G^{-1} \phi \rangle \nonumber \ee by admitting arbitrary positive
$G$, i.e. arbitrary propagator $G(x,y)$.
  [In the translation invariant case, the propagator is
considered as a function $G(x-y)$ of a single variable].
Diagrammatic expressions exhibit results in their
dependence on
$G$, as sums of products of propagators $G(x,y)$.
Therefore they
have a universal meaning, and the diagrammatic expansion
for the
\emph{same} quantity must agree, no matter how the
computation is
organized. Our way of organizing the expansion is more
economical
in that it avoids having to go through the moments; they
are here
summed from the start and $Z(q)$ is their generating
function. We
obtain diagrammatic expansions for (connected parts
$\langle
M^p\rangle^c $ of) $\langle M^p\rangle$ by expanding our
$q$-dependent propagator $\Gamma$ and $Tr \ln \Gamma$ in
powers of
$q$, cf. eqs.(\ref{Gamma}), (\ref{TrLnGamma}) and
(\ref{Gamma0}).
To recover the diagrammatic expansion of Ref. \cite{Bram3}
for
$\langle M^p\rangle / \langle M \rangle^p $ one must
divide by
$\langle M \rangle^p$. As shown in the Appendix, the loop
expansion can be
  reorganized so that $\langle M \rangle$ appears as a
factor multiplying $q$.
This implies that the diagrams for $\langle M^p\rangle /
\langle M \rangle^p $
are precisely the diagrams for $\langle M^p\rangle$ which
contain no tadpole.
A tadpole is a line in the graph whose source and target
agree; tadpoles produce factors $TG(0)/2 = Tg_1/2$.
Figure 2 shows the connected 2-loop graphs without
tadpoles
which contribute for $p=3,4$, in spin wave approximation.

\begin{figure}[h]
\centering 
\hspace{-.5 cm}
\includegraphics[width=.35\textwidth]{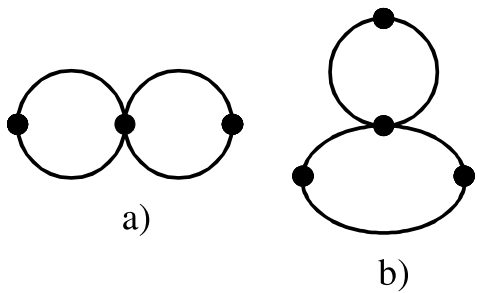}
\vspace{-.2cm} \caption{Connected 2-loop graphs in spin wave
approximation. a) contribution to $i^3\langle M^3 \rangle^c/3! =
(-iqT)^3\alpha_3$, b) contribution to $i^4\langle M^4 \rangle^c/4!
= (-iqT)^4\alpha_4$. } \label{graphs}
\end{figure}
For the sake of comparison let us evaluate these diagrams.

We remember that in a scalar field theory with interaction
$V=\frac \lambda {n!} \phi^n$, vertices have $n$ incident lines
and contribute a factor $-\lambda$. The $-$sign comes because the
Boltzmannian involves $e^{-V}$. In the expansion of $\frac
{iq}{N}\cos \phi_x$, even powers $\phi_x^n$ appear with factors
$1/n!$ and alternating sign. The Feynman rules for our theory in
harmonic approximation say that $iq\langle M \rangle + \ln Z(q)$
is the sum of all connected vacuum diagrams, and in these diagrams
there is
\begin{itemize}
\item a factor $\frac {iq}{N} (-)^{n/2}$ for every vertex with an even number $n$ of incident lines
\item a factor $TG(x-y)$ for a line from a vertex at $x$ to a vertex at $y$
\item a sum over $x$ for every vertex
\item a symmetry factor $1/g$, where $g$ is the number of elements of the symmetry group of the graph.
\end{itemize}
Since every vertex contributes a factor $q$, the vacuum diagrams
with $k$ vertices sum to the summand $(-iqT)^k\alpha_k(T)$ in the
expansion (\ref{expand}). Contributions with tadpoles can be
eliminated by setting $g_1=0$. The symmetry factor for the graph
(a) is $ \frac 1 8$, and the symmetry factor for the graph (b) is
$\frac 1 4 $. Using that $\sum_y G(x-y)G(y-x) = G^2(0)=Ng_2 $ etc,
the diagrams evaluate to results in agreement with
eq.(\ref{alpha2harm}), viz. \ba
(-iqT)^3 \alpha_3^{2harm} &=& (iq)^3\frac {T^4} {8} g_2^2 + O(g_1) \label{E.2}\\
(-iqT)^4 \alpha_4^{2harm} &=& -(iq)^4\frac {T^5} {4} g_2 g_3  + O(g_1) \label{E.3}\ .
\ea
We see that the contribution of the diagrams shown in Fig. \ref{graphs} does not vanish in the thermodynamic limit.

The result that $\langle M \rangle $ can be factored out confirms
that hyperscaling remains true to all orders in the loop
expansion. Hyperscaling asserts that $\langle M \rangle^2 /
\sigma^2 $ has a thermodynamic limit. Within the frame work of a
loop expansion, this must be true order by order in $T$. This is
not trivially true because $g_1$ has no thermodynamical limit.
Hyperscaling also remains true in the presence of anharmonic
corrections because these merely renormalize the temperature, as
we saw. Our 2-loop results for $\sigma$ and $\langle M \rangle $
which include anharmonic corrections give \be \sigma = \sqrt{\frac
{g_2} 2}\, T \langle M \rangle\left( 1 +\frac 1 {2} T +
O(T^2)\right). \ee for large $N$, which should be compared with
eq. (19) of Ref. \cite{Bram3}.

We thank the referee for asking questions which led to additions
and improvements to the paper. G. P. wants to thank for the kind
hospitality of the II. Institute for Theoretical Physics of the
University of Hamburg, where this work was completed. Partial
support provided by FONDECYT, Project No. 1020010 is gratefully
acknowledged.

\section*{Appendix: Normal ordered procedure}
We introduce the normalized Gausssian measure with covariance $TG$
\be d\mu_{TG}(\phi) = \frac 1 Z_0 \exp \left\{ -\frac \beta 2 \sum
|\nabla_\mu \phi_x|^2\right\} D\phi \ee It has the characteristic
function \be \int d\mu_{TG}(\phi) e^{i \langle j,\phi \rangle} =
e^{-\frac T 2 \langle j, G j \rangle } . \ee If $A(\phi)$ is a
function of $\phi $ which is integrable with respect to the
Gaussian measure, it can be written in normal ordered
form~\cite{GlimmJaffe}. The normal ordering operation is typically
indicated by ::, but it depends on the covariance of the Gaussian
measure.
\begin{eqnarray}
A(\phi) &=& \ : B(\phi): , \\
B(\phi) &=& \int d\mu_{TG}(\xi ) A(\phi + \xi) \ .
\end{eqnarray}
In particular
\be e^{i\alpha \phi_x} = e^{- \frac 1 2\alpha^2T G(0)}
:e^{i\alpha
\phi_x}: \ee
It follows that
\begin{eqnarray}
\cos \phi_x = e^{- \frac 1 2 T G(0)} :\cos \phi_x:
\label{A.1} \\
\int d\mu_{TG} (\phi ) :\cos \phi_x : = 1 \ . \label{A.2}
\end{eqnarray}

Under general conditions, $B$ is a holomorphic function of
$\phi$
which can be expanded into a power series. In particular

\be :\cos \phi_x : = 1 - \frac 1 2 :\phi_x^2: + \frac 1
{4!}:\phi_x^4: + \dots \label{A.3} \ee

It is well known that the normal products $:\phi^p_x:$ are
always
eigenmodes of the linearized renormalization group at the
Gaussian
fix point determined by $TG$. In two dimensions, with fix
point Hamiltonian $\beta \CH_{SW}=\frac 1 2 \beta \sum
|\nabla_\mu
\phi|^2 $ they all have dimension $0$.

Consider now the partition function

\be \tilde{Z}(q) = e^{iq\langle M\rangle }Z(q). \ee

It follows from eq.(\ref{3}) that the moments

\be \langle M^p \rangle = \left( -i \frac
{\partial}{\partial
q}\right)^p \tilde{Z}(q)|_{q=0}. \ee

We restrict our attention to the spin wave approximation,
replacing $\CH$ by $\CH_{SW}$. Inserting eq.(\ref{A.1})
into the
definition (\ref{4}) of $Z(q)$, we obtain

\be \tilde{Z}(q) = \int d\mu_{TG}(\phi) \exp\left\{\frac
{iq}{N}
e^{-\frac T 2 G(0)} \sum_x :\cos \phi_x : \right\} \ee

From this and eq.(\ref{A.2}) we recover the known result for the
magnetization in spin wave approximation,

\be \langle M \rangle = e^{- \frac T 2 G(0) } \ee

Therefore \be \tilde{Z}(q) = \hat{Z}(\langle M \rangle q)
\ee with
a new partition function

\be \hat{Z}(q) = \int d\mu_{TG}(\phi) \exp\left\{ \frac
{iq}{N}\sum_x :\cos \phi_x : \right\} , \ee and \be
\langle M^p
\rangle / \langle M \rangle^p = \left( -i \frac
{\partial}{\partial q}\right)^p \hat{Z}(q)|_{q=0} \ee

The partition function $\hat{Z}(q)$ is exactly like
$\tilde{Z}(q)$
except for the normal ordering of $\cos \phi_x$. In a
diagrammatic
expansion, normal ordering eliminates tadpoles. The
diagrams which
contribute to $\langle M^p \rangle / \langle M \rangle^p$
are
therefore exactly the diagrams without tadpoles which
contribute
to $\langle M^p \rangle $.

It is more convenient to consider truncated expectation
values \be
\langle M^p \rangle^c / \langle M \rangle^p = \left( -i
\frac
{\partial}{\partial q}\right)^p \ln \hat{Z}(q)|_{q=0}. \ee
They
have expansions in connected diagrams. The n-th order in
the loop
expansion for $\ln \hat{Z}$ sums connected diagrams with
$n$
loops. In 1-loop approximation only the term $ 1 - \frac 1
2:\phi_x^2: $ in the expansion (\ref{A.3}) of $:\cos
\phi_x:$
contributes, but the $\frac 1 {4!}:\phi_x^4:$-term
  begins to contribute in 2-loop order, and so on.


\begin{thebibliography}{99}
\bibitem{Wilson} K. Wilson, Phys. Rev. B \textbf{4}, 3174
(1971).

\bibitem{Bram1} S.T. Bramwell et al., Phys. Rev. Lett
\textbf{84}, 3744
(2000).

\bibitem{Bram2} S. T. Bramwell et al., Nature
\textbf{396}, 552 (1998).

\bibitem{BTW} P. Bak, C. Tang and K. Weisenfeld, Phys.
Rev. Lett. \textbf{59}%
, 381 (1997).

\bibitem{SNP} K. Sneppen, Phys. Rev. Lett. \textbf{69},
3538 (1992).

\bibitem{PSR} P. Sinha-Ray, L.Borda de Agua and H.J.
Jensen, Physica D
\textbf{157}, 186 (2001).

\bibitem{ECG} E. Caglioti, V. Loreto, H. Hermann and M.
Nicodemi, Phys. Rev.
Lett. \textbf{79}, 1575 (1997).

\bibitem{Golden} V. Aji and N. Goldenfeld Phys. Rev. Lett.
\textbf{86}, 1007
(2001).

\bibitem{Archam} P. Archambault et al., J. Appl. Phys.
\textbf{83}, 7234
(1998).

\bibitem{Palma1} G. Palma, T. Meyer and R. Labb\'e, Phys.
Rev. E \textbf{66}%
, 026108 (2002).

\bibitem{Bram3} S. T. Bramwell et al., Phys. Rev. E
\textbf{63}, 041106
(2001).

\bibitem{Palma2} G. Palma, Z. Phys. C \textbf{54}, 679
(1992)

\bibitem{Symanzik} K. Symanzik, J. Math. Phys. \textbf{7},
510 (1966)

\bibitem{GlimmJaffe} J. Glimm, A. Jaffe, {\em Quantum
Physics: A Functional Integral Point of View} Springer,
New York 1987
\end{thebibliography}
\end{document}